\documentclass{elsart3p}
\usepackage{times}
\usepackage[numbers,sort&compress]{natbib}
\usepackage{amssymb}
\usepackage[final]{graphicx}
\usepackage{units}

\newcommand{\dm}{\partial_\mu}

\begin{document}

\begin{frontmatter}
\title{The Standard Model Higgs boson as the inflaton}

\author[EPFL,INR]{Fedor Bezrukov},
\ead{Fedor.Bezrukov@epfl.ch}
\author[EPFL]{Mikhail Shaposhnikov}
\ead{Mikhail.Shaposhnikov@epfl.ch}

\address[EPFL]{
  Institut de Th\'eorie des Ph\'enom\`enes Physiques,
  \'Ecole Polytechnique F\'ed\'erale de Lausanne,
  CH-1015 Lausanne, Switzerland}
\address[INR]{
  Institute for Nuclear Research of Russian Academy of Sciences,
  Prospect 60-letiya Oktyabrya 7a,
  Moscow 117312, Russia}

\date{4 December 2007}

\begin{abstract}
  We argue that the Higgs boson of the Standard Model can lead to
  inflation and produce cosmological perturbations in accordance with
  observations.  An essential requirement is the non-minimal coupling
  of the Higgs scalar field to gravity; no new particle besides
  already present in the electroweak theory is required.
\end{abstract}

\begin{keyword}
  Inflation \sep Higgs field \sep Standard Model \sep Variable Planck
  mass \sep
  Non-minimal coupling
  \PACS 98.80.Cq \sep 14.80.Bn
\end{keyword}

\end{frontmatter}

\section{Introduction}

The fact that our universe is almost flat, homogeneous and isotropic
is often considered as a strong indication that the Standard Model
(SM) of elementary particles is not complete.  Indeed, these puzzles,
together with the problem of generation of (almost) scale invariant
spectrum of perturbations, necessary for structure formation, are most
elegantly solved by inflation
\cite{Starobinsky:1979ty,Starobinsky:1980te,Mukhanov:1981xt,Guth:1980zm,%
  Linde:1981mu,Albrecht:1982wi}.  The majority of present models of
inflation require an introduction of an additional scalar---the
``inflaton''.  This hypothetical particle may appear in a natural or
not so natural way in different extensions of the SM, involving Grand
Unified Theories (GUTs), supersymmetry, string theory, extra
dimensions, etc.  Inflaton properties are constrained by
the observations of fluctuations of the Cosmic Microwave Background
(CMB) and the matter distribution in the universe.  Though the mass
and the interaction of the inflaton with matter fields are not fixed,
the well known considerations prefer a heavy scalar field with a mass
$\sim \unit[10^{13}]{GeV}$ and extremely small self-interacting
quartic coupling constant $\lambda \sim 10^{-13}$ \cite{Linde:1983gd}.
This value of the mass is close to the GUT scale, which is often
considered as an argument in favour of existence of new physics
between the electroweak and Planck scales.

The aim of the present Letter is to demonstrate that the SM itself can
give rise to inflation.  The spectral index and the amplitude of
tensor perturbations can be predicted and be used to distinguish this
possibility from other models for inflation; these parameters for the
SM fall within the $1\sigma$ confidence contours of the WMAP-3
observations \cite{Spergel:2006hy}.

To explain our main idea, consider Lagrangian of
the SM non-minimally coupled to gravity,
\begin{equation}
\label{main}
  L_{\mathrm{tot}}= L_{\mathrm{SM}} - \frac{M^2}{2} R -\xi H^\dagger HR
  \;,
\end{equation}
where $L_{\mathrm{SM}}$ is the SM part, $M$ is some mass parameter,
$R$ is the scalar curvature, $H$ is the Higgs field, and $\xi$ is an
unknown constant to be fixed later.\footnote{In our notations the
  conformal coupling is $\xi=-1/6$.}  The third term in (\ref{main})
is in fact required by the renormalization properties of the scalar
field in a curved space-time background \cite{Birrell:1982ix}.  If
$\xi=0$, the coupling of the Higgs field to gravity is said to be
``minimal''.  Then $M$ can be identified with Planck scale $M_P$
related to the Newton's constant as $M_P=(8\pi
G_N)^{-1/2}=\unit[2.4\times 10^{18}]{GeV}$.  This model has ``good''
particle physics phenomenology but gives ``bad'' inflation since the
self-coupling of the Higgs field is too large and matter fluctuations
are many orders of magnitude larger than those observed.  Another
extreme is to put $M$ to zero and consider the ``induced'' gravity
\cite{Zee:1978wi,Smolin:1979uz,Spokoiny:1984bd,Fakir:1990iu,Salopek:1988qh},
in which the electroweak symmetry
breaking generates the Planck mass
\cite{vanderBij:1993hx,CervantesCota:1995tz,Bij1995}. This happens if
$\sqrt{\xi}\sim1/(\sqrt{G_N} M_W)\sim10^{17}$, where $M_W\sim\unit[100]{GeV}$
is the electroweak scale.  This model may give ``good'' inflation
\cite{Spokoiny:1984bd,Fakir:1990iu,Salopek:1988qh,Kaiser:1994wj,%
  Kaiser:1994vs,Komatsu:1999mt} even if the scalar self-coupling is of
the order of one, but most probably fails to describe particle physics
experiments.  Indeed, the Higgs field in this case almost completely
decouples from other fields of the SM\footnote{This can be seen most
  easily by rewriting the Lagrangian (\ref{main}), given in the Jordan
  frame, to the Einstein frame, see also below.}
\cite{vanderBij:1993hx,CervantesCota:1995tz,Bij1995}, which corresponds
formally to the infinite Higgs mass $m_H$.  This is in conflict with
the precision tests of the electroweak theory which tell that $m_H$
must be below $\unit[285]{GeV}$ \cite{:2005ema} or even
\unit[200]{GeV} \cite{PDG2007} if less conservative point of view is
taken.

These arguments indicate that there may exist some intermediate choice
of $M$ and $\xi$ which is ``good'' for particle physics and for
inflation at the same time.  Indeed, if the parameter $\xi$ is
sufficiently small, $\sqrt{\xi} \lll 10^{17}$, we are very far from the
regime of induced gravity and the low energy limit of the theory
(\ref{main}) is just the SM with the usual Higgs boson.  At the same
time, if $\xi$ is sufficiently large, $\xi \gg 1$, the scalar field
behaviour, relevant for chaotic inflation scenario
\cite{Linde:1983gd}, drastically changes, and successful inflation
becomes possible.  We should note, that models of chaotic inflation
with both nonzero $M$ and $\xi$ were considered in literature
\cite{Spokoiny:1984bd,Futamase:1987ua,Salopek:1988qh,Fakir1990,Kaiser:1994vs,%
  Libanov1998,Komatsu:1999mt}, but in the context of either GUT or
with an additional inflaton having nothing to do with the Higgs field
of the Standard Model.

The Letter is organised as follows.  We start from discussion of
inflation in the model, and use the slow-roll approximation to find
the perturbation spectra parameters.  Then we will argue in Section
\ref{sec:radcorr} that quantum corrections are unlikely to spoil the
classical analysis we used in Section \ref{sec:cmb}.  We conclude in
Section~\ref{sec:concl}.

\section{Inflation and CMB fluctuations}
\label{sec:cmb}

Let us consider the scalar sector of the Standard Model, coupled to
gravity in a non-minimal way. We will use the unitary gauge
$H=h/\sqrt{2}$ and neglect all gauge interactions for the time being,
they will be discussed  later in Section \ref{sec:radcorr}.  Then the
Lagrangian has the form:
\begin{equation}
  \label{eq:1}
  \begin{array}{l@{\,}l}
    \displaystyle
    S_{J} =\int d^4x \sqrt{-g} \Bigg\{&\displaystyle
    - \frac{M^2+\xi h^2}{2}R
    \\
    &\displaystyle
    + \frac{\dm h\partial^\mu h}{2}
    -\frac{\lambda}{4}\left(h^2-v^2\right)^2
    \Bigg\}
    \;.
  \end{array}
\end{equation}
This Lagrangian has been studied in detail in many papers on inflation
\cite{Salopek:1988qh,Fakir1990,Kaiser:1994vs,Komatsu:1999mt}, we will
reproduce here the main results of
\cite{Salopek:1988qh,Kaiser:1994vs}.  To simplify the formulae, we
will consider only $\xi$ in the region $1\ll\sqrt{\xi}\lll10^{17}$, in
which $M \simeq M_P$ with very good accuracy.

\begin{figure}
  \centering
  \includegraphics[width=\columnwidth]{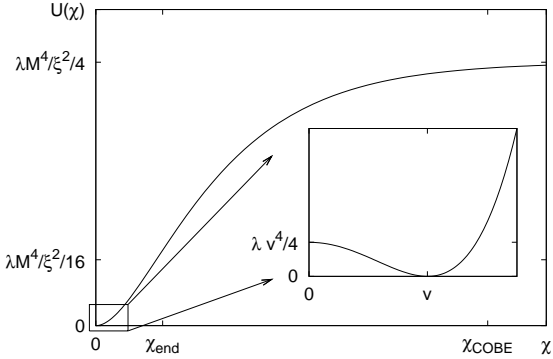}
  \caption{Effective potential in the Einstein frame.}
  \label{fig:Ueff}
\end{figure}

It is possible to get rid of the non-minimal coupling to gravity by
making the conformal transformation from the Jordan frame to the
Einstein frame
\begin{equation}
  \label{eq:2}
  \hat{g}_{\mu\nu} = \Omega^2 g_{\mu\nu}
  \;,\quad
  \Omega^2 = 1 + \frac{\xi h^2}{M_P^2}
  \;.
\end{equation}
This transformation leads to a non-minimal kinetic term for the Higgs
field. So, it is convenient to make the change to the new scalar field
$\chi$ with
\begin{equation}
  \label{eq:3}
  \frac{d\chi}{dh}=\sqrt{\frac{\Omega^2+6\xi^2h^2/M_P^2}{\Omega^4}}
  \;.
\end{equation}
Finally, the action in the Einstein frame is
\begin{equation}
  \label{eq:4}
    S_E =\int d^4x\sqrt{-\hat{g}} \Bigg\{
    - \frac{M_P^2}{2}\hat{R}
    + \frac{\dm \chi\partial^\mu \chi}{2}
    - U(\chi)
    \Bigg\}
    \;,
\end{equation}
where $\hat{R}$ is calculated using the metric $\hat{g}_{\mu\nu}$ and
the potential is
\begin{equation}
  \label{eq:5}
  U(\chi) =
  \frac{1}{\Omega(\chi)^4}\frac{\lambda}{4}\left(h(\chi)^2-v^2\right)^2
  \;.
\end{equation}
For small field values $h\simeq\chi$ and $\Omega^2\simeq1$, so the
potential for the field $\chi$ is the same as that for the initial
Higgs field.  However, for large values of $h\gg M_P/\sqrt{\xi}$ (or
$\chi\gg\sqrt{6}M_P$) the situation changes a lot.  In this limit
\begin{equation}\label{eq:hlarge}
  h\simeq \frac{M_P}{\sqrt{\xi}}\exp\left(\frac{\chi}{\sqrt{6}M_P}\right)
  \;.
\end{equation}
This means that the
potential for the Higgs field is exponentially flat and has the form
\begin{equation}
  \label{eq:6}
  U(\chi) = \frac{\lambda M_P^4}{4\xi^2}
  \left(
    1+\exp\left(
      -\frac{2\chi}{\sqrt{6}M_P}
    \right)
  \right)^{-2}
  \;.
\end{equation}
The full effective potential in the Einstein frame is presented in
Fig.~\ref{fig:Ueff}.  It is the flatness of the potential at $\chi\gg
M_P$ which makes the successful (chaotic) inflation possible.

\begin{figure}
  \centering
  \includegraphics[width=\columnwidth]{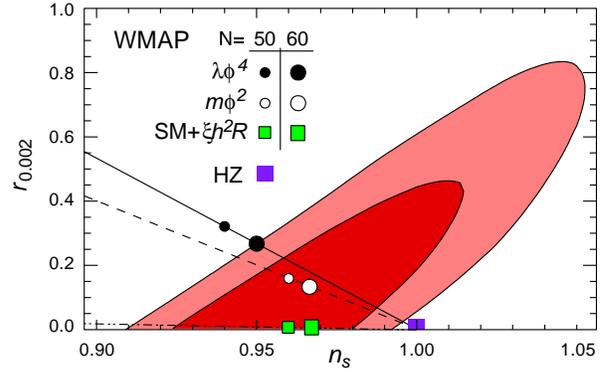}
  \caption{The allowed WMAP region for inflationary parameters ($r$,
    $n$).  The green boxes are our predictions supposing 50 and 60
    e-foldings of inflation.  Black and white dots are predictions of
    usual chaotic inflation with $\lambda\phi^4$ and $m^2\phi^2$
    potentials, HZ is the Harrison-Zeldovich spectrum.}
  \label{fig:wmap}
\end{figure}

Analysis of the inflation in the Einstein frame\footnote{The same
  results can be obtained in the Jordan frame
  \cite{Makino1991,Fakir:1992cg}.} can be performed in standard way
using the slow-roll approximation.  The slow roll parameters (in
notations of \cite{Lyth:1998xn}) can be expressed analytically as
functions of the field $h(\chi)$ using (\ref{eq:3}) and (\ref{eq:5})
(in the limit of $h^2\gg M_P^2/\xi\gg v^2$),
\begin{eqnarray}
  \label{eq:7}
  \epsilon & =& \frac{M_P^2}{2}\left(\frac{dU/d\chi}{U}\right)^2
  \simeq\frac{4 M_P^4 }{3
   \xi^2h^4}
  \;, \\
  \eta & = & M_P^2\frac{d^2U/d\chi^2}{U}
  \simeq -\frac{4 M_P^2}{3 \xi h^2 }
  \;, \\
  \zeta^2 &= & M_P^4\frac{(d^3U/d\chi^3)dU/d\chi}{U^2}
  \simeq\frac{16 M_P^4 }{9\xi^2
   h^4}
  \;.
\end{eqnarray}
Slow roll ends when $\epsilon\simeq1$, so the field value at the end of
inflation is
$h_{\mathrm{end}}\simeq(4/3)^{1/4}M_P/\sqrt{\xi}\simeq1.07M_P/\sqrt{\xi}$.
The number of e-foldings for the change of the field $h$ from $h_0$ to
$h_{\mathrm{end}}$ is given by
\begin{equation}
  \label{eq:8}
  N = \int_{h_{\mathrm{end}}}^{h_0}
  \frac{1}{M_P^2}\frac{U}{dU/dh}\left(\frac{d\chi}{dh}\right)^2dh
  \simeq \frac{6}{8}\frac{h_0^2-h_{\mathrm{end}}^2}{M_P^2/\xi}
  \;.
\end{equation}
We see that for all values of $\sqrt{\xi}\lll10^{17}$ the scale
of the Standard Model $v$ does not enter in the formulae, so the
inflationary physics is independent on it.  Since interactions of the
Higgs boson with the particles of the SM after the end of inflation
are strong, the reheating happens right after the slow-roll, and
$T_{\mathrm{reh}}\simeq(\frac{2\lambda}{\pi^2
  g^*})^{1/4}M_P/\sqrt{\xi}\simeq\unit[2\times10^{15}]{GeV}$, where
$g^*=106.75$ is the number of degrees of freedom of the SM.  So, the
number of e-foldings for the the COBE scale entering the horizon
$N_{\mathrm{COBE}}\simeq62$ (see \cite{Lyth:1998xn}) and
$h_{\mathrm{COBE}}\simeq9.4M_P/\sqrt{\xi}$.  Inserting (\ref{eq:8})
into the COBE normalization $U/\epsilon=(0.027M_P)^4$ we find the
required value for $\xi$
\begin{equation}
  \label{eq:9}
  \xi \simeq \sqrt{\frac{\lambda}{3}}\frac{N_{\mathrm{COBE}}}{0.027^2}
  \simeq  49000\sqrt{\lambda}
  =  49000\frac{m_H}{\sqrt{2}v}
  \;.
\end{equation}
Note, that if one could deduce $\xi$ from some fundamental theory this
relation would provide a connection between the Higgs mass and the
amplitude of primordial perturbations.
The spectral index $n=1-6\epsilon+2\eta$ calculated for $N=60$
(corresponding to the scale $k=0.002/\mathrm{Mpc}$) is
$n\simeq1-8(4N+9)/(4N+3)^2\simeq0.97$.  The tensor to scalar
perturbation ratio \cite{Spergel:2006hy} is
$r=16\epsilon\simeq192/(4N+3)^2\simeq0.0033$.
The predicted values are well within one sigma of the current WMAP
measurements \cite{Spergel:2006hy}, see Fig.~\ref{fig:wmap}.

\section{Radiative corrections}
\label{sec:radcorr}

An essential point for inflation is the flatness of the scalar
potential in the region of the field values $h\sim10M_P/\sqrt{\xi}$,
what corresponds to the Einstein frame field
$\chi\sim 6 M_P$. 
 It is important that radiative corrections do not spoil
this property.  Of course, any discussion of quantum corrections is
flawed by the non-renormalizable character of gravity, so the
arguments we present below are not rigorous.

There are two qualitatively different type of corrections one can
think about.  The first one is related to the quantum gravity
contribution.  It is conceivable to think \cite{Linde:1987yb} that
these terms are proportional to the energy density of the field $\chi$
rather than its value and are of the order of magnitude $U(\chi)/M_P^4
\sim \lambda/\xi^2$.
They are small at large $\xi$ required by
observations.  Moreover, adding non-renormalizable operators
$h^{4+2n}/M_P^{2n}$ to the Lagrangian (\ref{eq:1}) also does not
change the flatness of the potential in the inflationary region.\footnote{Actually, in
  the Jordan frame, we expect that higher-dimensional operators are
  suppressed by the effective Planck scale $M_P^2+\xi h^2$.}

Other type of corrections is induced by the fields of the Standard
Model coupled to the Higgs field.  In one loop approximation these
contributions have the structure
\begin{equation}
\Delta U \sim \frac{m^4(\chi)}{64\pi^2} \log\frac{m^2(\chi)}{\mu^2}~,
\label{1loop}
\end{equation}
where $m(\chi)$ is the mass of the particle (vector boson, fermion, or
the Higgs field itself) in the background of field $\chi$, and $\mu$ is
the normalization point.  Note that the terms of the type $m^2(\chi)
M_P^2$
(related to quadratic divergences) do not appear in
scale-invariant subtraction schemes that are based, for example, on
dimensional regularisation (see a relevant discussion in
\cite{Shaposhnikov:2006xi,Meissner:2006zh,Shaposhnikov:2007nj,Meissner:2007xv}).
The masses of the SM fields can be readily computed
\cite{Salopek:1988qh} and have the form
\begin{equation}
  m_{\psi,A}(\chi) = \frac{m(v)}{v}\frac{h(\chi)}{\Omega(\chi)}
  \;,\quad
  m^2_H(\chi) = \frac{d^2U}{d\chi^2}
\end{equation}
for fermions, vector bosons and 
the Higgs (inflaton) field.  It is crucial that for large $\chi$
these masses approach different constants (i.e.\ the one-loop contribution
is as flat as the tree potential) and that (\ref{1loop}) is suppressed
by the gauge or Yukawa couplings in comparison with the tree term.  In
other words, one-loop radiative corrections do not spoil the flatness
of the potential as well.  This argument is identical to the one given
in \cite{Salopek:1988qh}.

Another important correction is connected with running of the
non-minimal coupling $\xi$ to gravity.  The
corresponding renormalization group equation is \cite{Buchbinder1992,Yoon1997}
\begin{equation}\label{eq:xi}
  \mu\frac{d\xi}{d\mu}=\left(\xi+\frac{1}{6}\right)\frac{\left(
    12\lambda+12y_t^2-\frac{9}{2}g^2-\frac{3}{2}{g'}^2
  \right)}{16\pi^2}\;,
\end{equation}
where $y_t=m_t/v$ is the top Yukawa coupling, $g$ and $g'$ are SU(2)
and U(1)
couplings of the Standard Model and $\mu$ is the characteristic
scale.  The renormalization of $\xi$ from $\mu\sim M_W$ to the Planck
scale is considerable, $\xi(M_P)\approx 2\xi(M_W)$.  At the
same time, the change of $\xi$ in the inflationary region is small,
$\delta\xi/\xi\approx0.2$.  Thus, the logarithmic running of $\xi$
does not change the behaviour of the potential required for inflation.

There is also the induced one-loop pure gravitational term of the form
$\xi^2 R^2/64\pi^2$.  During the inflationary epoch it is smaller
than the tree term $M_P^2R$ by the Higgs self-coupling $\lambda/64\pi^2$ and does not
change the conclusion.

\section{Conclusions}
\label{sec:concl}

In this Letter we argued that inflation can be a natural consequence of
the Standard Model, rather than an indication of its weakness. The
price to pay is very modest---a non-minimal coupling of the Higgs
field to gravity.  An interesting consequence of this hypothesis is
that the amplitude of scalar perturbations is proportional to the
square of the Higgs mass (at fixed $\xi$), revealing a non-trivial
connection between electroweak symmetry breaking and the structure of
the universe.  The specific prediction of the inflationary parameters
(spectral index and tensor-to-scalar ratio) can distinguish it from
other models (based, e.g.\ on inflaton with quadratic potential),
provided these parameters are determined with better accuracy.

The inflation mechanism we discussed has in fact a general character
and can be used in many extensions of the SM. Thus, the $\nu$MSM of
\cite{Asaka:2005an,Asaka:2005pn} (SM plus three light fermionic
singlets) can explain simultaneously neutrino masses, dark matter,
baryon asymmetry of the universe and inflation without introducing any
additional particles (the $\nu$MSM with the inflaton was considered in
\cite{Shaposhnikov:2006xi}). This provides an extra argument in favour
of absence of a new energy scale between the electroweak and Planck
scales, advocated in \cite{Shaposhnikov:2007nj}.

\section*{Acknowledgements}

The authors thank S. Sibiryakov, V. Rubakov, I. Tkachev, O.
Ruchayskiy, H.D. Kim, P. Tinyakov, and A. Boyarsky for valuable
discussions.  This work was supported by the Swiss National Science
Foundation.


\end{document}